\documentclass[a4paper,10pt]{article}

\usepackage{amssymb}
\usepackage{amsmath}
\usepackage[usenames,dvipsnames]{color}
\usepackage{hyperref}
\usepackage[left=.8in,right=.8in,top=.8in,bottom=.8in]{geometry}                  

\hypersetup{
    colorlinks=true,         
    linkcolor=blue,          
    citecolor=red,        
    urlcolor=Violet             
}

\newtheorem{theo}{Theorem}

\newcommand{\mean}[1]{\ensuremath{\lf\langle #1 \rt\rangle }}

\newcommand{\order}[1]{\ensuremath{\mathcal{O}(#1)}}

\def\be{\begin{equation}}
\def\ee{\end{equation}}
\def\bea{\begin{eqnarray}}
\def\eea{\end{eqnarray}}

\def\lf {\ensuremath{\left}}
\def\rt {\ensuremath{\right}}

\title{A Birkhoff theorem for Shape Dynamics}
\author{\bf Henrique Gomes\footnote{\href{mailto:gomes.ha@gmail.com}{gomes.ha@gmail.com}}\\\it  Department of Physics,  University of California, Davis,   CA, 95616}

\begin{document}

\maketitle

\begin{abstract}
Shape Dynamics is a theory of gravity that replaces refoliation invariance for spatial Weyl invariance. Those solutions of the Einstein equations that have global, constant mean curvature slicings, are mirrored by solutions in Shape Dynamics. However,   there are solutions of  Shape Dynamics that have no counterpart in General relativity,  just as there are solutions of GR that are not completely foliable by global constant mean curvature slicings (such as the Schwarzschild spacetime). 
It is therefore interesting  to analyze directly the equations of motion of Shape Dynamics in order to find its own solutions, irrespective of properties of known solutions of GR. Here I perform a first study in this direction by utilizing the equations of motion of Shape Dynamics in a spherically symmetric, asymptotically flat ansatz to derive an analogue of the Birkhoff theorem. There are two significant differences with respect to the usual Birkhoff theorem in GR. The first 
 regards the construction of the solution:  the spatial Weyl gauge freedom of shape dynamics is used to simplify the problem, and  boundary conditions are required. In fact the derivation is  simpler than the usual Birkhof theorem as no Christoffel symbols are needed. The second, and most important difference is that the solution obtained is uniquely the isotropic wormhole solution, in which no singularity is present, as opposed to maximally extended Schwarzschild. This provides an explicit example of the breaking of the duality between General relativity and Shape Dynamics, and exhibits some of its consequences.  \end{abstract}

\section{Introduction}
\subsection{Shape Dynamics}

Shape Dynamics is a theory of gravity,  formulated in the Hamiltonian 3+1 formalism. It possesses two dynamical propagating degrees of freedom, and has as kinematical variables the same $g_{ab}$ and $\pi^{ab}$ as the Hamiltonian version of ADM General Relativity. What is noteworthy about this model is the fact that it possesses spatial Weyl invariance, acting within each spatial surface on both the metric and on the momenta. It maintains the correct number of degrees of freedom since it does not have refoliation invariance. \footnote{To be more precise, it does not possess the corresponding phase space symmetry, since in phase space  refoliations per se are not meaningful.}  

Shape Dynamics is intimately related to specific gauge-fixings of Hamiltonian 3+1 ADM. It is a theory that takes as its geometric observables \emph{spatial} conformal--diffeomorphism invariants, as opposed to \emph{space-time} diffeomorphism invariants.  
 The gauge-fixings of ADM that are related to Shape Dynamics  are either constant mean curvature (CMC) for the closed spatial manifold case,  or  maximal slicing for the open manifold case. To be more precise, suppose that I restrict  spatial manifolds $\Sigma$ (to be identified with the spatial part of globally hyperbolic spacetimes,  $\Sigma\times \mathbb{R}$) to be open. Then Shape Dynamics is a gauge theory whose reduced phase space is intimately connected to a maximal slicing $g_{ab}\pi^{ab}=\pi=0$ gauge fixing of ADM.  There is a special property of these gauge-fixings which is the reason they are related to Shape Dynamics. That property is that these gauge-fixings also moonlight as generators of spatial Weyl transformations. In the same way that $\pi^{ij}_{~;j}= 0$ generates spatial diffeomorphisms in Hamiltonian ADM general relativity, $\pi= 0$ (maximal slicing) generates spatial Weyl transformations, while $\pi-\mean{\pi}\sqrt{g}=0$ (CMC) generates total-volume preserving conformal transformations \cite{CS+V, SD_first}.
 
 I should also stress the well-known fact that the conformally reduced ADM gravity  - which is equivalent to imposing  maximal slicing gauge as a second class constraint - is much richer than simply looking  for maximal slicings of given space-time solutions of GR. It is a reduced phase space theory with its own symplectic structure and equations of motion, its own dynamics, which we see as primary. From solution curves in reduced phase space variables -- let's call them $(\rho_{ab}(t), \sigma^{ab}(t))$ -- one then has an algorithm to reconstruct a line element on a 4-dimensional manifold $g_{\mu\nu}(\rho(t), \sigma(t), \xi(t))$, where $\xi(t)$ is a choice of shift vector field along the solution.  In the same way that some space-time solutions, such as Schwarzschild, do not possess complete foliations by maximal slices (and thus no representation in reduced phase space) the reduced phase space equations have solutions that possess obstructions to having a space-time built up from them using the above algorithm. Degeneracy of the reconstructed 4-metric is such an obstruction, and the one we will encounter. However, such obstructions cannot be seen  in the phase space solution $(g_{ab}(t),\pi^{ab}(t)$)  of the equations of motion of Shape Dynamics.  
 
 On the other hand, if a given Einstein space-time has a complete maximal slicing it will be also a solution of the gauge-fixed reduced theory. The point of the matter is that any solution of the  maximal slicing gauge-fixed (conformally reduced) theory has a counterpart in Shape Dynamics, which means it has a spatial Weyl-invariant representation.  Thus from the point of view of Shape Dynamics, a global space-time satisfying Einstein's equation is physical only if it possesses  a global maximal slicing.\footnote{For a compact closed spatial manifold the criterium gets enlarged to encompass constant mean curvature foliations.} 
 
The breaking of the duality between the two theories is represented on the reconstructed line element by a collapse of a particular  space-time interval between the hypersurfaces, the interval required to propagate the maximal slicing condition. However, \emph{this collapse need not be physical in either of the two theories}. It clearly is not a physical statement in GR, where for example the lapse indicates merely a failure of the maximal slicing  gauge. Analogously, in shape dynamics it signals a failure of the ``space-time" gauge.  In both sides, the collapse itself is a gauge-dependent statement. Nonetheless, the \emph{translation} between the two theories, which have different gauge symmetries available to explore, \emph{does} break down when the lapse collapses, and the two theories can have different continuations from then on.  There are solutions of Shape Dynamics which are only locally isomorphic to space-time solutions of General Relativity: they do not form space-times, but are by definition still  physical in the Shape Dynamics sense. 

I will find such a solution, and furthermore show that it is the unique such solution if I assume asymptotic flatness and spherical symmetry in vacuum. This imposition of boundary conditions already signals a difference between the Shape Dynamics analogue of the Birkhoff theorem and the GR one, which yields a static 4-metric from spherical symmetry irrespective of the boundary conditions. Further differences lie at the construction level, since I used in the construction the Shape Dynamics equations of motion \emph{and Weyl gauge freedom}. Using these methods it turns out that the construction is in fact much simpler than the usual Birkhoff theorem (there is no need to calculate Christoffel symbols for instance).\footnote{Some months after the initial completion of this work, it was called to my attention that Deser \cite{Deser_Sch} used basically the same method to derive the isotropic line element of the exterior Schwarzschild solution. In that work, he splits the metric into a conformal factor and a ``conformal geometry" part, which effectively implies working on the conformally reduced phase space. What I assume here that is also implicit there (and not necessary in the Birkhof theorem) are the fixed boundary conditions.} Finally, the way in which the actual spherically symmetric solution for Shape Dynamics differs from that of general relativity is that it builds an isotropic wormhole solution (which is \emph{not} a vacuum solution of GR) as opposed to (maximally extended) Schwarzschild. However, it matches the exterior Schwarzschild solution. This mismatch is possible because the isotropic solution, when taken for the full radial domain, is not \emph{ spacetime} -- continuous on the horizon. However, it comes from a  perfectly continuous curve in phase space representing star--collapse in shape dynamics. 

\subsection{Birkhoff's theorem in GR}
In General Relativity, Birkhoff's theorem \cite{Birkhoff}\footnote{In fact, the theorem today known as Birkhoff's theorem was first discovered by Jebsen in \cite{Jebsen}. } states that a spherically symmetric solution of the Einstein field equations in vacuum $R_{\mu\nu}=0$, must contain an extra Killing vector field. The resulting solution is given by the Schwarzchild metric: 
\be\label{equ:Schwarzschild}ds^2= -\frac{1}{1-\frac{2m}{r}}dt^2+(1-\frac{2m}{r})dr^2+r^2(d\theta^2+\sin^2(\theta)d\phi^2)
\ee
where the domain of the variables are $-\infty<t<\infty$ , $0<r<\infty$, $0<\theta<\pi$ and $-\pi<\phi<\pi$. If the Killing vector field is time-like, the resulting metric will be locally static and asymptotically flat, corresponding to the region $r>2m$ in \eqref{equ:Schwarzschild}.  If it is space-like, it will be homogeneous and it will run into a singularity at $r=0$, corresponding to the region $r<2m$ in \eqref{equ:Schwarzschild}. 

Clearly something goes wrong at $r=2m$ in \eqref{equ:Schwarzschild}: the radial component $g_{rr}$ of the metric blows up, while $g_{tt}$ vanishes. However, this is a \emph{coordinate singularity}, since no physical observable, such as invariants  of the curvature, blows up in that region.

\subsubsection*{Coordinate invariance and the degeneracy of the 4-metric}

 To characterize some object as coordinate invariant we must first define what we take to be valid coordinate transformations. As usual, I define it as a smooth diffeomorphism: a smooth transformation with a smooth inverse. The requirement of existence of the inverse implies that to be \emph{characterized as a coordinate transformation the transformation's  Jacobian  has to have a finite, non-zero determinant.} A coordinate change which obeys these conditions is used to obtain a maximal extension of Schwarzchild (Kruskal–-Szekeres)  which avoids the pathologies of the original coordinate system employed in \eqref{equ:Schwarzschild}.  For future purposes I call attention to the fact that the \emph{determinant} of the 4-metric of the Schwarzschild solution above \eqref{equ:Schwarzschild} $\det{g_{\mu\nu}}=r^2\sin^2\theta$ \emph{does not vanish anywhere in the coordinate domain}  (although the $g_{00}$ component vanishes at the horizon). Unlike what occurs in the above case however,  a  4-metric which is degenerate in some region (i.e. whose determinant vanishes in that region) cannot be transformed into a regular (non-degenerate) 4-metric by a viable coordinate transformation. The degeneracy of the metric is equivalent to a physical statement (physical in the context of general relativity): the 4-volume element collapses at the given set. This distinction will become important in order to show that the Shape Dynamics solution \emph{is not} a space-time solution.

\section{Shape Dynamics for asymptotically flat conditions}

\subsection{Construction}

The construction of Shape Dynamics for the case of a closed spatial manifold is more complicated than the one I am  about to present here \cite{SD_first}.
Since the construction of the theory for the case where the underlying spatial manifold has boundary is much simpler, I will present it here without reference to the changes that have to be made in the closed manifold case. 

The first step is to write out the constraints of canonical GR in its 3+1 ADM form: 
\begin{eqnarray}
\label{equ:scalar constraint}S(x):= \frac{G_{abcd}\pi^{ab}\pi^{cd}}{\sqrt g}(x)-R(x)\sqrt g(x)=0\\
\label{equ:momentum constraint} H_a:={\pi^{a}_b}_{;a}=0
\end{eqnarray}
where the points $x$ belong to an open 3-manifold $\Sigma$, $g_{ab}$ is the spatial 3-metric and its conjugate momenta $\pi^{ab}$. The scalar constraint \eqref{equ:scalar constraint} generates on-shell refoliations of spacetime, while the momentum constraint generates foliation preserving diffeomorphisms. The second step is to extend phase space in a trivial manner, including the variables $\phi$ and its canonically conjugate momenta $\pi_\phi$. This entails the appearance of third set of first class constraints 
$$\pi_\phi=0
$$
Now we perform a canonical transformation in the extended phase space with coordinates $(g_{ab},\pi^{ab},\phi, \pi_\phi)$ of the form: 
$$t_\phi:(g_{ab},\pi^{ab},\phi,\pi_\phi)\mapsto (e^{4\phi}g_{ab},e^{-4\phi}\pi^{ab},\phi,\pi_\phi-4\pi)$$
where $\pi=g_{ab}\pi^{ab}$. An extra first class constraint in this extended theory arises:  $\pi_\phi-4\pi\approx 0$. 

The smeared scalar constraint \eqref{equ:scalar constraint} becomes, for $\phi=\ln\Omega$  
\be\label{LY}t_\psi S(N)=\int_\Sigma \left( \nabla^2\Omega+\frac{1}{8}R\Omega-\frac{1}{8}\pi^{ab}\pi_{ab}\Omega^{-7}\right)N\approx 0 
\ee
Ignoring boundary terms (see \cite{Asymptotic_SD} for details on how to treat the boundary terms), the smeared diffeomorphism constraint becomes
\be\label{equ:conf_diffeo_constraint}
t_\phi H_a(\xi^a)=\int_\Sigma \left(\pi^{ab}\mathcal{L}_\xi g_{ab}+\pi_\phi\mathcal{L}_\xi \phi \right) d^3 x\approx 0
\ee
while we now have a new constraint which was not present in ADM: 
\be\label{equ:conformal} C(\rho)=\int_\Sigma \left(\pi_\phi-4\pi\right)\rho\approx 0
\ee
Where $N$,  $\xi^a$ and $\rho$ are Lagrange multipliers.

As is usual for dynamical systems on manifolds that have a boundary, we must add terms to the total Hamiltonian to ensure differentiablity and thus the well-posedness of the equations of motion. 
Let us summarize here some of the assumptions and results of the boundary treatment, given in  \cite{Asymptotic_SD}.  First, the boundary conditions taken for the asymptotically flat metric variables and Lagrange multipliers of the scalar and diffeomorphism constraint  are:
\begin{eqnarray}
g_{ab}\rightarrow \delta_{ab}+\order{r^{-1}}~~&,& ~~\pi^{ab}\rightarrow \order{r^{-2}}\label{equ:fall-off}\\
N\rightarrow 1+\order{r^{-1}}~~&,& ~~\xi^a\rightarrow \order{1}\nonumber
\end{eqnarray}
 The fall-off condition for the remaining Lagrange multiplier $\rho$ can be determined from 
the requirements of consistency and finiteness of the boundary terms. 
 In \cite{Asymptotic_SD} I have shown that in order to  preserve the asymptotic form of the metric variables and obtain finite boundary terms, the asymptotic behavior of the Lagrange multiplier and the conformal  variables should be  $\rho\rightarrow \order{r^{-1}}$,  and  $e^\phi\rightarrow 1+ \order{r^{-1}}$ and $\pi_\phi\rightarrow \order{r^{-2}}$.  

To recover the ADM dynamical system, one must merely gauge-fix this extended theory by setting $\phi=0$. 
For the construction of Shape Dynamics, we perform the gauge-fixing $\pi_\phi=0$ on the extended system. The only constraint that does not commute with this gauge-fixing is exactly \eqref{LY}. This constraint can be solved for $\Omega$ \cite{York} in terms of the variables $g_{ab}$ and $\pi^{ab}$, and thus together with the setting of $\pi_\phi=0$  the system is reduced, and the Dirac bracket also reduces to the canonical Poisson brackets of the variables $(g_{ab}, \pi^{ab})$. It turns out that in the present case there remains a total Hamiltonian residing on the boundary $\partial\Sigma$. It is of the form: 
\be\label{equ:SD_H}
H_{\mbox{\tiny{SD}}}[g,\pi]=-\int_{\partial\Sigma}d^2y \sqrt h(2 (k-k_o)+8r_e \Omega_o^{,e})
\ee 
where $k$ is the trace of the extrinsic curvature of the boundary, $k_o$ is the extrinsic curvature of the boundary if embedded in flat space,  $\Omega_o$ is the functional of $g_{ab}, \pi^{ab}$ which solves \eqref{LY}, $h_{ab}$ is the metric at the boundary and $r^a$ is the normal to the boundary. The other remaining constraints are the usual spatial diffeomorphism constraints 
$$\int d^3 x\left(\pi^{ab}\mathcal{L}_\xi g_{ab}\right)=0
$$
with $\xi^a$ now respecting the asymptotically flat boundary conditions. Finally, from the reduction of the constraint $C=\pi_\phi - 4\pi$ we also obtain a Weyl (or conformal) constraint $\pi=0$ (whose smearing $\rho$ also must satisfy the appropriate boundary conditions).

\subsection{Equations of motion}
 Shape Dynamics is  obtained from a Linking theory (what I have called  so far the extended theory), where all quantities are local. Thus the simplest way to  formulate Shape Dynamics' equations of motion, boundary charges, counter-terms and fall-off conditions is to consider these in the larger setting of the Linking theory, and then use phase space reduction.


As shown in the appendix, the most straightforward way of calculating the equations of motion of shape dynamics requires first finding a solution to the lapse fixing equation:\footnote{Note that here I do not require the solution to be positive everywhere.}
\be\label{LFE} e^{-4\phi_o}(\nabla^2 N_o+2g^{ab}\phi^o_{,a}N^o_{,b}) -N_oe^{-6\phi_o} G_{abcd}\pi^{ab}\pi^{cd}=0
\ee
where I have denoted the solution of \eqref{LY} by $\Omega_o[g,\pi]=e^{\phi_o[g,\pi]}$. 
The equations valid in the present case are:
\begin{eqnarray}
\label{equ:eom_g}\dot g_{ab}&=&4\rho g_{ab}+2e^{-6\phi_o}\frac{N_o}{\sqrt g}\sigma_{ab}+\mathcal{L}_{\xi}g_{ab}\\
\dot\sigma^{ab}&=& N_oe^{2\phi_o}\sqrt{g}\left(R^{ab}-2\phi_o^{;ab}+4\phi_o^{,a}\phi_o^{,b}-\frac{1}{2}R g^{ab}+2 \nabla^2\phi_o g^{ab}\right)\nonumber\\
&~&-\frac{N_o}{\sqrt g}e^{-6\phi_o}\left(2(\sigma^{ac}\sigma^b_c)-\frac{1}{2}(\sigma^{cd}\sigma_{cd})g^{ab}\right)\nonumber\\
&~&-e^{2\phi_o}\sqrt {g}\left(N_o^{;ab}-4\phi_o^{(,a}N_o^{,b)}-\nabla^2N_o g^{ab}\right)+\mathcal{L}_{\xi}\sigma^{ab}-4\rho\sigma^{ab}\label{equ:eom_pi}
\end{eqnarray}
Here I have already eliminated the trace of the momentum from the equations, writing the traceless momenta as $\sigma^{ab}:=\pi^{ab}-\frac{1}{3}\pi g^{ab}$. This is valid on-shell, but is not necessary for the well-posedness of the equations. Note also the presence of the conformal gauge terms $4\rho g_{ab}$ and  $-4\rho\sigma^{ab}$. 
 To obtain a solution of ADM, I must impose the gauge-fixing $\phi_o[g,\pi]=0$, which is a condition of course on $g_{ab}$ and $\pi^{ab}$. 

\section{ Birkhoff's theorem in Shape Dynamics}\label{sec:Birkhoff}

In this section I derive the main results of the paper. The precise statement of the theorem that I will derive is the following: 
\begin{theo}[A Birkhoff-type theorem for shape dynamics]~

Given the asymptotically flat conditions \eqref{equ:fall-off} and spherical symmetry, the  only solution (up to spatial conformal transformations) of equations \eqref{LFE}, \eqref{LY}, \eqref{equ:eom_g} and \eqref{equ:eom_pi} are given, in spherical coordinates by: 
\be g_{ab}=\eta_{ab}~, ~~ \pi^{ab}=0~, ~~\Omega_o= (1+\frac{m}{2r})~, ~~N=\left(\frac{1-\frac{m}{2r}}{1+\frac{m}{2r}}\right)
\ee
where $\eta_{ab}$ is the Euclidean flat metric, $m$ is the shape dynamics mass of the solution, given in \eqref{equ:SD_H}, and $0<r<\infty$. The family of line elements that are reconstructed are thus: 
\be\label{ds} ds^2= -\left(\frac{1-\frac{m}{2r}}{1+\frac{m}{2r}}\right)^2dt^2+ (1+\frac{m}{2r})^4\left(dr^2+r^2(d\theta^2+\sin^2(\theta)d\varphi^2)\right)
\ee
which represents an Einstein-Rosen bridge.
\end{theo}

To show the theorem, one has, under the conditions specified above, four equations to solve: equations \eqref{LY} and \eqref{LFE}, which determine the conformal factor and the lapse on a single spatial hypersurface, and equations \eqref{equ:eom_g} and \eqref{equ:eom_pi} which are the equations of motion for the dynamical variables that are not Lagrange multipliers. 
The  strategy for the calculation is as follows: 
\begin{enumerate}
\item  Spherical symmetry implies the 3-geometry is conformally flat,
\item Equation \eqref{equ:eom_g} implies that, on the initial spatial hypersurface at $t=0$, for  a particular initial value for the gauge field  $\rho$, the initial momenta and the shift vector vanish.
\item  Equation \eqref{LY} with mass charge $m$ is solved for $\Omega$.  
\item  Equation \eqref{LFE} is solved for $N$ using the solution to $\Omega$. It depends on an integration constant $b$.
\item Equation \eqref{equ:eom_pi} is simplified using the previous steps,  and it is shown that unless $b=0$ the equations of motion for $\pi^{ab}$ do not preserve the assumed boundary conditions. This implies that $\dot \pi^{ab} = 0$ at $t = 0$ and thus $\pi^{ab} = 0$ at all times.
\item Having zero momenta,  $\dot g_{ab}$ is reduced to a pure-gauge evolution, so the conformal 3-geometry is static, and with $N, \Omega, \xi, g_{ab} ,\pi^{ab}$ that have been calculated one reconstructs the line element \eqref{ds}.

\end{enumerate}


\subsection*{Construction}

The objective criterion to identifying conformally flat metrics is through the use of the Cotton tensor. Much like the Weyl tensor in 4-dimensions, the Cotton tensor encodes the conformally invariant degrees of freedom of the 3-metric $g_{ij}$. 
As is easy to verify, spherical symmetry implies that  the Cotton tensor for the 3-metric vanishes, thus there is no coupling to the Weyl-invariant degrees of freedom, except through the boundary conditions.

 Thus our metric is conformally flat, and we can restrict ourselves to a phase space in which $g_{ab}=\Omega^4(r,t)\eta_{ab}$ where $\eta_{ab}=\mbox{diag}(1, r^2, r^2\sin^2(\theta))$ in the spherical coordinates $\{r,\theta, \varphi\}$. In \eqref{equ:eom_g}, I can replace $g_{ab}$ by $\eta_{ab}$,  $\phi$ by $\ln\Omega$, and $\dot{g_{ab}}=4\Omega^3\dot\Omega \eta_{ab}$ on the lhs. To see that this can be done,  rewriting $g_{ab}$ as $\Omega^4(r,t) \eta_{ab}$ in \eqref{LY}, \eqref{LFE} \eqref{equ:eom_g}, \eqref{equ:eom_pi} then due to the conformal covariance of these equations they can be rewritten as the same equations  with $\eta_{ab}$ in place of $g_{ab}$, and with $\phi+\ln\Omega$ in place of $\phi$. But $\phi+\ln\Omega$ is just a redefinition of $\phi$ and I can call $\phi' = \phi+\ln\Omega$  and solve for $\phi'$. This concludes the first step in the proof. 

For the second step, I first rewrite equation \eqref{equ:eom_g}: 
\be \label{equ:eom_g2}\dot \Omega^4\eta_{ab}=4\rho \eta_{ab}+2e^{-6\phi_o}\frac{N_o}{\sqrt \eta}\sigma_{ab}+\mathcal{L}_{\xi}(\Omega^4\delta_{ab})\ee In accordance with spherical symmetry, for simplicity I assume that the shift is of the the form $\xi^a=\xi(r,t)\delta^a_r$. The Lie derivative of the metric with respect to a vector field of this form is: 
\be \mathcal{L}_{\vec{\xi}}(\tilde\Omega\eta_{ab})=\xi\tilde\Omega_{,r}\eta_{ab} -2r\tilde\Omega\xi(\delta^\theta_a\delta^\theta_b+\sin^2(\theta)\delta^a_\varphi\delta^b_\varphi)
+2\tilde\Omega\xi_{,r}\delta^r_a\delta^r_b
\ee
where for convenient notation I wrote $\tilde\Omega=\Omega^4$. I can rewrite this in a form which has more terms proportional to the metric:
$$ 2r\tilde\Omega\xi(\delta^\theta_a\delta^\theta_b+\sin^2(\theta)\delta^a_\varphi\delta^b_\varphi)= 2\eta_{ab} \frac{\tilde\Omega\xi}{r}-2\frac{\tilde\Omega\xi}{r}\delta^r_a\delta^r_b
$$
Now putting all the terms proportional to the metric on the lhs of \eqref{equ:eom_g2} I get: 
\be\label{equ:new dot g} (f(r,t)-\rho)\eta_{ab}= \gamma(r,t)\sigma_{ab}+ 2\tilde\Omega\delta^r_a\delta^r_b(\xi_{,r}-\frac{\xi}{r})
\ee where $f(r,t)=\dot{\tilde\Omega}+\frac{\tilde\Omega\xi}{r}\sim \mathcal{O}(r^{-1})$ , and 
$\gamma(r,t)$ is an unimportant collection of the prefactors of the trace-free part of the momenta,  $\sigma^{ab}$.

I choose the initial conformal gauge freedom to satisfy $\rho(r,0)=f(r,0)$. Then the lhs of \eqref{equ:new dot g}  is identically zero,  and taking the trace of the equation I obtain $\xi_{,r}-\frac{\xi}{r}=0$, which means $\xi=\alpha r$ and, to obey the boundary conditions on $\xi^a$, we must  have $\alpha=0=\xi$. Taking then the traceless part of equation \eqref{equ:new dot g} then more importantly implies that the momenta initially vanishes: ${\sigma_{ab}}_{|t=0}=0$. This concludes step 2. 

   The vacuum  equation \eqref{LY} now gets simplified to, 
  $$\partial^a\partial_a \Omega=0$$ 
   with the general boundary conditions $1+\order{1/r}$ the solution is given by
$ \Omega=1+\frac{m(t)}{2r}$, where I have kept an arbitrary time dependence on the (spatial) constant $m$. \footnote{The solution $ \Omega=1+\frac{m}{2r}$ cannot in fact be vacuum everywhere, it requires a singular contribution at the origin, which plays the role of the source. If we input a point source energy density $p$ at $r=0$,  in radial coordinates the scalar equation \eqref{LY} 
 becomes
$ \partial^2\Omega= 2\pi p\delta(r)\Omega(0)^5$,
where the $2\pi$ factor is exactly what appears from the rhs of the Hamiltonian constraint ($2\pi$ comes from $16\pi/8$, where the $8$ is the factor for $\nabla^2\Omega$ in \eqref{LY}) and where $m=p\Omega(0)^5$. Note that $\Omega(0)$ is a divergent quantity, requiring a regularization between $m$ and $p$.} 
   To see that the boundary conditions on the next to leading order contribution to $\Omega$ are justified, I calculate the asymptotic energy, given by \eqref{equ:SD_H}. Trivially, one obtains   $\partial_r\phi\rightarrow-\frac{m}{2r^2}$, and asymptotically $k\rightarrow \partial_j g_{ij}-\partial_i(\delta^{kl}g_{kl})$ which is equal to zero since $g_{ij}=\delta_{ij}$. Thus we are left with (after restoration of the $\frac{1}{16\pi}$ factor):
\be H_{\mbox{\tiny{SD}}}=-\frac{1}{2\pi}\int_{\partial\Sigma}d^2y\partial_r\phi (r^2\sin({\theta})d\theta d\phi)=m(t)
\ee
The first class property of the constraints then imply that $\dot{m}=\{H_{\mbox{\tiny{SD}}},H_{\mbox{\tiny{SD}}}\}=0$ which ensures the conservation of energy.
This finishes the third step (note that I can  now safely set $f(r,0)=\rho(r,0)=0$ in step 2).

Moving on,  equation  \eqref{LFE} becomes:
$$\partial^a\partial_a N+2(1+\frac{m}{2r})_{,r}N_{,r}=0
$$
 with asymptotically flat boundary conditions, the solution is given by $N=1-\frac{2b}{m+2r}$, where $b$ is an integration constant.  If I input the lapse $N=1-\frac{2b}{m+2r}$ and $\phi=\ln{(1+\frac{m}{2r})}$ into \eqref{equ:eom_pi}, I obtain
\be\label{equ:lapse_determination}\dot\pi^{ab}=(1+\frac{m}{2r})^2
\left(
\begin{array}{ccc}
 \frac{8 (-1+b) m}{r (m+2 r)^2} & 0 & 0 \\
 0 & -\frac{4 (-1+b) m r \text{sin}^2(\theta )}{(m+2 r)^2} & 0 \\
 0 & 0 & -\frac{4 (-1+b) m r}{(m+2 r)^2} \\
\end{array}
\right)
\ee
We see that to maintain the boundary conditions on the metric momenta, $\pi^{ab}\sim \order{r^{-2}}$  we must have $b=1$, thus $\dot\pi^{ab}_{t=0}=0$, which implies fact $\pi^{ab}(t)=0$. Thus,  I find the solution  compatible with my boundary conditions  by taking $b=1$. This completes steps 4 and 5. 

To reconstruct the 4-metric from this, we must gauge-fix $\phi_o=0$, or in other words $\Omega_o=1$. I use the Weyl invariance of $e^{4\phi_o[g,\pi]}g_{ab}$, $e^{-4\phi_o[g,\pi]}\pi^{ab}$ and go to the gauge where $\Omega_o=1$.  Rewriting the lapse solution as $1-\frac{2m}{m+2r}=\frac{1-\frac{m}{2r}}{1+\frac{m}{2r}}$, the reconstructed line element, \emph{here valid for the full range $0<r<\infty$}, is:
\be \label{equ:SD_new}
ds^2= -\left(\frac{1-\frac{m}{2r}}{1+\frac{m}{2r}}\right)^2dt^2+ (1+\frac{m}{2r})^4\left(dr^2+r^2(d\theta^2+\sin^2(\theta)d\varphi^2)\right)
\ee
Note that I make a careful distinction here between line elements and spacetimes. A line element can be degenerate and have other pathologies when seen from the spacetime view. Equation \eqref{equ:SD_new} completes the last step and the theorem. $\square$

\section{Conclusions}
I start the conclusions by pointing out that in the construction of the solution, unlike what is done in the usual Birkhoff theorem,   both the boundary conditions and the ability to absorb the initial slice pure gauge terms in $\rho$ are utilized. I also note that usual proofs of Birkhoff's theorem do not utilize explicitly the equation of motion $\dot\pi^{ab}$. This equation determined a leftover integration constant $b=1$ in the lapse \eqref{LY} from \eqref{equ:lapse_determination} and allowed us to recover exactly the isotropic black hole line element.  I also note that the usual Birkhof theorem  requires one to calculate Christoffel symbols and curvature tensors, calculations which we can completely avoid in the present context. In \cite{Deser_Sch}, Deser uses very similar methods to derive the isotropic line element of the \emph{exterior} Schwarzschild solution. However, he interprets the solution merely as a novel way to derive the \emph{exterior} Schwarzschild solution (since he there already assumes the spacetime is static). Here I do not make such an assumption and yet am able to derive a full analogue of the Birkhoff theorem using the equations of motion.

The first thing to notice regarding the line element solution itself is that interpreting it as a vacuum \emph{space-time}, one obtains that the local 4-volume collapses at the horizon: the 4-metric becomes degenerate, and thus \emph{does not possess an inverse}.\footnote{One should not confuse this \emph{space-time} degeneracy of the metric with the usual degeneracy of a Killing horizon in GR, which is topologically a 3-surface but metrically a 2-surface. For  Killing horizons in GR, the \emph{space-time} metric is still \emph{non degenerate} at the horizon. } The degeneracy of the metric is an invariant statement about the metric under  space-time diffeomorphisms, i.e. it is a \emph{physical statement}. This solution can not be considered as a vacuum space-time per se, but  instead what one would obtain if attempting to describe a valid Shape Dynamics solution in space-time language. As we will see shortly, the solution represents a wormhole, and it is well known that in order to have solutions which possess an Einstein-Rosen bridge   requires in general relativity a singular shell of (exotic) matter at the horizon  \cite{Visser_book}.   
In other words, without the inclusion of a singular shell of matter at $r=m/2$, the line element \eqref{equ:SD_new} is a  GR solution \emph{only for $r>m/2$}.
However,  for Shape Dynamics it \emph{should} be considered as a physical vacuum solution, since no observables of Shape Dynamics (spatial conformal diffeomorphism invariant quantities) break down at the horizon, and during collapse the trajectory in phase space also suffers no obstruction \cite{Tim_collapse}.

 Furthermore,  at the no back-reaction level, we can in principle use the ``reconstructed line element" as a background for the motion of particles in Shape Dynamics. This entails that we can calculate most things that we would normally calculate in GR, keeping in mind that Lorentz-invariance - which is not a \emph{fundamental} symmetry of Shape Dynamics - is broken at $r=m/2$. A more dynamical and appropriate way to calculate evolution in this background is to couple e.g. scalar fields to the Hamiltonian and solve the augmented lapse fixing equations \eqref{LFE} and Lichnerowicz-York equation \eqref{LY}. Koslowski \cite{Tim_collapse}  has performed the necessary calculations for a thin-shell collapse. The main ingredient there is to notice that due to spherical symmetry, the conformal geometry itself does not couple to the source fields, and thus remains fixed throughout evolution. The path in phase space is parametrized by a single parameter denoting the radius of the shell, and for shape dynamics only the continuity of this path in phase space is required.

One might think that the the region $r=m/2$ would prohibit any time-like trajectory from going into the black hole. However, one can in fact calculate certain properties, such as the proper time of a radially infalling geodesic merely form symmetry and energy conservation arguments. For the metric \eqref{equ:SD_new}, this was done originally by Poplawski, in \cite{Poplawski:2009aa}, where he shows that:
\begin{itemize}
\item
An observer in the asymptotically flat region will not see any difference between this  solution and a Schwarzschild black hole. A simple calculation shows that the infalling radial geodesic takes infinite proper time to reach $r=0$. \emph{This is of course a physical, observable distinction with respect to Schwarzschild.}
\item  The infalling observer will (after finite proper time) experience falling into a mirror universe.    The mirror universe is obtained by replacing the isotropic radial coordinate with $r\to m^2/(4r)$, which can be checked to leave the form of the line element (\ref{equ:SD_new}) invariant. Thus the solution has the property of \emph{inversion}, associated to conformal invariance (as for instance in the method of images in electrodynamics).\footnote{This type of duality is of great avail in String theory on e.g. a compactified torus, toggling between strong and weak coupling regimes \cite{Polchinski_book}. }  
\item  Notice that the mirror universe essentially decompactifies the region near $r=0$ by a conformal transformation that puts infinite spatial distance between $r=0$ and any other point in a given $t=const.$ surface. The corresponding Penrose diagram does not contain a singularity, it consists of two identical causal diamonds, spatially bound by $r=0, r=\infty$ and separated by the horizon at $r=m/2$.\footnote{Note however that in this model Lorentz invariance does not hold at the horizon, which might complicate the standard interpretation of a Penrose diagram.} 
\item One might worry about Penrose's singularity theorem. However, there is a discontinuity of the expansion scalar at the event
horizon of the wormhole, which  prevents the expansion scalar from decreasing to $-\infty$, as it
would occur inside a Schwarzschild black hole. \end{itemize}
After performing these calculations, Poplawski warns the reader that:   
\begin{quote}``While the Schwarzschild metric
is the spherically symmetric solution to the Einstein field equations in vacuum if we solve these equations using
the Schwarzschild coordinates, the Einstein-Rosen bridge is the spherically symmetric solution to the Einstein field
equations if we solve these equations using isotropic coordinates for a source which is vacuum everywhere \emph{except at
the surface $r=m/2$. It has been shown that the Einstein-Rosen bridge metric  is not a solution of the vacuum
Einstein equations but it requires the presence of a nonzero energy-momentum tensor source $T_{\mu\nu}$ that is divergent and
violates the energy conditions at the throat of the wormhole}.\cite{Visser_book} " [Our italics].\end{quote}
This is the issue that shape dynamics does not possess, in our case \eqref{equ:SD_new} \emph{is} a vacuum solution \emph{for the full range of the radial variable $r$}. 

 The natural follow-up question is if this singularity-avoiding property persists in all solutions of Shape Dynamics. {In this respect, I should mention that space-times in maximal slicing have a well-known singularity avoidance property for its Eulerian observers (which are the natural observers to take in the Shape Dynamics side) \cite{3+1_book}.} Indeed, after a first version of this paper an axi-symmetric solution with  largely the same qualitative properties obtained here was found \cite{SD_Kerr}.

It is interesting to speculate on some possible implications of the results of this paper. For example, suppose that  isolated   particles in Shape Dynamics could be modeled by the solution \eqref{equ:SD_new}.\footnote{This is a terrible idea for any physical elementary particle. The issue there is that the ratio between  the electric charge and the mass of an electron $e/m$ don't obey the bounds of Reissner-Nordstrom solutions (and thus would contain a naked singularity). An analogous objection holds if one considers spins and the ratio between the angular momentum and the mass $J/m$. } After traversing the Schwarzschild radius, gravitational interaction becomes weaker, not stronger, presenting thus a classical anti-screening effect.\footnote{There is a sense in which one can define such a concept as classical (anti) screening. Classical counter-parts of renormalization group properties have been recently applied to a variety of models (\cite{Dvali} and references therein). }. One other possible speculation is if this has any relevance in the context of ER=EPR resolutions of the firewall scenario, where Einstein-Rosen bridges are identified with entangled regions of space-times \cite{Cool_horizons}.

\subsection*{Comments after first version}
Some months after the initial completion of this work, it was called to my attention that Deser \cite{Deser_Sch} used very similar methods to derive the isotropic line element of the exterior Schwarzschild solution. 

\section*{Acknowledgements}

The author would like to thank  Tim Koslowski for ongoing conversations on this topic. I would also like to thank Steve Carlip for reading the manuscript and suggesting several important improvements. HG was supported in part by the U.S.
Department of Energy under grant DE-FG02-91ER40674.

\appendix

\section{Deriving the equations of motion: the closed manifold case}

I will start the derivation with the closed spatial manifold case, since that is the setting in which shape dynamics was originally constructed. In that case, the Weyl transformations do not have fast decay properties at asymptotic infinity of course, but are instead required to preserve the total volume of space.  
Before we start, consider the following. Suppose $\Delta$ is a given differential operator with a one-dimensional kernel, generated by the function $N_o(x)$. In this case the inverse operator $\Delta^{-1}$ yields only:
\be\label{equ:local_inverse_Delta2} \int d^3 x'\Delta^{-1}(x,x')\Delta(x',y)=\delta(x,y)-a N_o(x)\sqrt g(y)
\ee
where $a$ is an arbitrary constant.\footnote{I should mention that in the closed manifold case I assumed a normalization condition on $N_o$, namely that its space integral be unity. }

\subsection{Dirac brackets}

Given this, let's see explicitly what happens with the Dirac brackets. 
For the Dirac brackets as usually defined we have 
\be\label{equ:Dirac_closed}
 \{\cdot, \cdot\}_{\mbox{\tiny{DB}}}=\{\cdot ,\cdot\}-
 \{~\cdot~,\chi_A\}\Delta_{AB}^{-1}\{\chi_B,~\cdot~\}\ee where $A$ and $B$ denote both the continuous variables as well as the different sets of constraints, and I use summation (including integration). Furthermore, formally we have $\Delta_{AB}^{-1}\Delta^{AC}=\delta^C_B$. 
  The following identity should be valid: 
\be\label{equ:usual_DB}\{\cdot_{|\chi=0}, \cdot_{|\chi=0}\}_{\mbox{\tiny{DB}}}=\{\cdot, \cdot\}_{\mbox{\tiny{DB}}|\chi=0}
\ee

Although to define the constraints that will enter into the inverse Dirac matrix one takes weak equalities, one should \emph{not} take these equalities when calculating the bracket. In fact, one needs to take the inverse of the full second class matrix of constraints. In the generic case of a block diagonal matrix of operators of the form:
\be \label{Dirac_matrix}
D^{AB}=\left(\begin{array}{cc}
A& \Delta \\
-\Delta & 0\\ 
\end{array}\right)
\ee
where $\Delta$ is an invertible operator, the inverse is given by:
\be\label{correct_Dirac}
D^{-1}_{AB}=\left(\begin{array}{cc}
0& -\Delta^{-1} \\
\Delta^{-1} & \Delta^{-1}A\Delta^{-1} \\ 
\end{array}\right)
\ee

The second class constraints are included in $t_\phi S$ and $\pi_\phi$, whose Poisson bracket I will write as the differential operator $\Delta(x,y)=\{t_\phi S(x), \pi_\phi(y)\}$. As mentioned before, this operator has a non-trivial kernel. In the case of closed manifolds this kernel is one-dimensional. 

For the set of constraints $t_\phi S(x), \pi_\phi(x)$, since $\pi_\phi(x)$ form a set of abelian constraints, the full Poisson bracket matrix is exactly of the form \eqref{Dirac_matrix}, with  $A= \{t_\phi S,t_\phi S\}$.
Using \eqref{correct_Dirac}, the Dirac brackets are then given by: 
 \be\label{equ:Dirac_closed2}
 \{\cdot, \cdot\}_{\mbox{\tiny{DB}}}=\{\cdot ,\cdot\}-
 \{~\cdot~,\pi_\phi\}\Delta^{-1}\{{ t_{\phi}S},~\cdot~\}-\{~\cdot~,{ t_{\phi}S}\}\Delta^{-1}\{\pi_\phi ,~\cdot~\}-\{~\cdot~,\pi_\phi\}(\Delta^{-1}A\Delta^{-1})\{\pi_\phi ,~\cdot~\}
  \ee
 The bracket property $ \{\chi, \cdot\}_{\mbox{\tiny{DB}}}=0$ clearly holds, and thus we have the projectability condition:
 \be \{\cdot_{|\chi=0}, \cdot_{|\chi=0}\}_{\mbox{\tiny{DB}}}=\{\cdot, \cdot\}_{\mbox{\tiny{DB}}|\chi=0}\ee
Further, since $\pi_\phi=0$ is one of the second class constraints appearing in the  Dirac bracket, and I am eliminating $\phi$, for the reduced phase space variables, the  Dirac brackets obey:
\be\label{full_DBs} \{\cdot_{|\chi=0}, \cdot_{|\chi=0}\}_{\mbox{\tiny{DB}}}= \{\cdot_{|\chi=0}, \cdot_{|\chi=0}\}=\{\cdot, \cdot\}_{\mbox{\tiny{DB}}|\chi=0}\ee

\subsubsection*{Equations of motion} 
 Let us here just check some of the assertions made for the equations of motion.
 {For the alternative construction mentioned above I have $\{\pi_\phi(y), t_{\phi}S(N)\}=\Delta N(y)$, where $\Delta$ has a kernel. I obtain:
\begin{eqnarray*}
\{g_{ab}(x), t_{\phi}S(N) \}_{\mbox{\tiny{DB}}}&=&\{g_{ab}(x), t_{\phi}S(N) \}-\int d^3 x'd^3 y\{g_{ab}(x), t_{\phi}S(x')\}\Delta^{-1}(x',y)\Delta N(y)\\
 ~&=& \{g_{ab}(x), t_{\phi}S(N) \}- \{g_{ab}(x), t_{\phi}S(N) \} + a(\int d^3 x'\sqrt g N)\int d 3 y N_o(y)\{g_{ab}(x), t_{\phi}S(y) \}\\
 ~&=&  a (\int \sqrt g d^3 x' N)\int d^3 y\{g_{ab}, { t_{\phi}S}(y)\}N_o(y)
\end{eqnarray*} 
where now $\phi$ is taken to solve the full LY equation (but is still restricted to be total volume preserving). } 
  This means we can  calculate $\{g_{ab}, t_{\phi}S(N) \}$ and then, after the calculations are performed, input the substitution $\phi\mapsto \phi_o$ and $N\mapsto N_o$ arising from reduction. That is   
\be \{g_{ab}, t_{\phi}S(N) \}_{{\mbox{\tiny{DB}}}|\pi_\phi=\phi-\phi_o=0}=\{g_{ab}, t_{\phi_o}S(N_o)\}= (\int \sqrt g d^3 x' N)\int d^3 y\{g_{ab}, { t_{\phi}S}(y)\}_{|\phi-\phi_o=0}N_o(y)\ee
The same of course occurs with the $\pi^{ab}$ equations of motion.

\section{The open manifold case}

The advantage of using the kernel in the Green's functions can be seen directly here. The only difference from \eqref{equ:Dirac_closed2} is that I use a term $t_{\phi}S+t_{\phi}B$ instead of $t_{\phi}S$:
\begin{align}\label{equ:Dirac_open}
~& \{\cdot, \cdot\}_{\mbox{\tiny{DB}}}=\{\cdot ,\cdot\}-
 \{~\cdot~,\pi_\phi\}\Delta^{-1}\{{ t_{\phi}S-t_\phi B},~\cdot~\}\nonumber\\
 &-\{~\cdot~,{ t_{\phi}S-t_\phi B}\}\Delta^{-1}\{\pi_\phi ,~\cdot~\}-\{~\cdot~,\pi_\phi\}(\Delta^{-1}A\Delta^{-1})\{\pi_\phi ,~\cdot~\}
  \end{align}
where $A= \{{ t_{\phi}S-t_\phi B}, { t_{\phi}S-t_\phi B}\}$, where I have added the regulating boundary term: 
$$t_\phi B(x)= \int_{\partial\Sigma} t_\phi B(x)\delta(x,y)\sqrt h(y) d^2 y $$

The main question we have to address to fully define \eqref{equ:Dirac_open} is then to identify what we will take to be the kernel of $\Delta$. By the properties of $\Delta$, this is of course uniquely determined by the boundary conditions (for functions of compact support the kernel is trivial).

Suppose for instance we take $M_i$ to be a function obeying the boundary conditions $i$, for which the only non-trivial solution for the lapse fixing equation is given by $N_i$. This implies then, according to the same logic as before:
\be
\label{open_Delta}\Delta^{-1}\Delta M_i= M_i- N_i
\ee
Note that under reduction the only thing we have is that
$$ t_\phi S(x)+t_\phi B(x)\rightarrow t_{\phi_o} B(x)
$$
and I have thus fully accounted for the lapse solutions to the LFE appearing through the action of the Dirac brackets.  

\subsubsection*{Equations of motion}
For an arbitrary lapse function $M_i$ obeying the same asymptotics as the solution of the lapse fixing equation \eqref{LFE}, $N_i$, I have
\be
\{g_{ab}, t_{\phi}S(M_i) \}_{\mbox{\tiny{DB}}}=a(M_i)\int d^3 x\{ g_{ab}, t_\phi(S+B)(x) \}N_i(x)
\ee
the same holding with the momenta. These yield the equations of motion \eqref{equ:eom_g} and \eqref{equ:eom_pi}.


\end{document}